\documentclass[preprint]{aastex}
\begin{document}

\title{THE DETECTION OF COLD DUST IN CAS A: EVIDENCE FOR THE FORMATION OF METALLIC NEEDLES IN THE EJECTA}

\author{Eli Dwek}
\affil{Laboratory for Astronomy and Solar Physics \\ NASA Goddard Space Flight Center,
Greenbelt, MD 20771, \\ e-mail: eli.dwek@nasa.gov}

\begin{abstract}
Ejecta from core collapse supernovae contain a few solar masses of refractory elements and can therefore be the most important source of interstellar dust if these elements condense efficiently into solids. However, infrared observations of young supernova remnants, such as Cas~A or Kepler, and observations of SN~1987A have detected only $\sim 10^{-3}$~M$_{\odot}$ of hot dust in these objects. Recently, Dunne et al. (2003) obtained 450 and 850~$\mu$m SCUBA images of Cas~A, and reported the detection of 2--4~M$_{\odot}$ of cold, 18~K, dust in the remnant. Here we show that their interpretation of the observations faces serious difficulties. Their inferred dust mass ignores the effect of grain destruction by sputtering, and is larger than the mass of refractory material in the ejecta of a 10 to 30~M$_{\odot}$ star. The cold dust model faces even more difficulties if the 170~$\mu$m observations of the remnant are included in the analysis, decreasing the cold dust temperature to $\sim$~8~K, and increasing its mass to $\gtrsim$ 20 M$_{\sun}$.
We offer here a more plausible interpretation of their observation, in which the cold dust emission is generated by conducting needles in the ejecta. 
The needle properties are completely determined by the combined submillimeter and X-ray observations of the remnant.
The needles are collisionally heated by the shocked gas. They are very efficient emitters at submillimeter wavelengths, and with a resistivity of a few $\mu \Omega$~cm can readily attain a temperature of 8~K. Taking the destruction of needles into account, a dust mass of only $10^{-4}$ to $10^{-3}$~M$_{\odot}$ is needed to account for the observed SCUBA emission. The needles consist of metallic whiskers with $\lesssim$ 1\% of embedded impurities, that may have condensed out of blobs of material that were expelled at high velocities from the inner metal--rich layers of the star in an asymmetric explosion. Conductive needles may also be the source of the cold dust emission detected by Morgan et al. (2003) in Kepler. Aligned in the magnetic field, needles may give rise to observable polarized emission. The detection of submillimeter polarization will therefore offer definitive proof for a needle origin for the cold dust emission. Supernovae may still be proven to be important sources of interstellar dust, but the evidence is still inconclusive. 
\end{abstract}

\section{INTRODUCTION}
Cas A, the young remnant of a Type II supernova (SN), has been suggested as an ideal observing target for dust that may have formed in the explosive ejecta (Dwek \& Werner 1981). Initial ground-based searches for dust  (Wright et al. 1980, Dinerstein et al. 1982) yielded negative results. Dust was first detected in Cas~A in the {\it Infrared Astronomical Satellite} ({\it IRAS}) all--sky survey (Mezger et al. 1986, Dwek et al. 1987, Braun 1987, Arendt 1989, Saken, Fesen, \& Shull 1992). However, due to the low spatial resolution of the {\it IRAS} images, its origin whether interstellar, circumstellar or supernova, could not be ascertained. 
Higher resolution infrared images of Cas A, obtained with the {\it Infrared Space Observatory} ({\it ISO}) satellite, revealed the presence of hot dust in the fast--moving knots of the remnant, thereby unambiguously establishing its  SN origin (Lagage et al. 1996, Arendt, Dwek, \& Moseley 1999, Tuffs et al. 1999). Dust temperatures were typically $\sim$ 100--200~K, consistent with those expected for shock--heated dust. Dust masses were between $10^{-7} - 10^{-3}$~M$_{\sun}$, depending on the size of the observed region, significantly below the mass of condensable elements expected to be present in the ejecta of a massive progenitor star (Woosley \& Weaver 1995). 

Supernovae are potentially the most important source of interstellar dust and they play an important role in establishing the elemental depletion pattern in the interstellar medium (Dwek \& Scalo 1980, Dwek 1998, Jones 2000). Furthermore, the early rise in the UV-optical opacity of high redshift galaxies suggests copious dust production in these objects (Todini \& Ferrara 2001). The lack of any observational evidence for the formation of massive amounts of dust in supernovae is therefore quite frustrating. In light of this, the recent detections of emission from cold dust in Cas~A (Dunne et al. 2003) and Kepler (Morgan et al. 2003) are very encouraging, since they may provide the first observational confirmation of the role of supernovae as prominent interstellar dust sources. Their findings require therefore careful scrutiny. In our analysis we will concentrate on the cold dust detection from Cas~A.

The 450 and 850 $\mu$m SCUBA images of Cas~A  (Dunne et al. 2003) revealed a new, cold, dust component, primarily concentrated in the southern half of the remnant, and bounded between the forward and reverse shocks, located at angular radii of, respectively, 153~\arcsec and 95~\arcsec (Gotthelf et al. 2001). The total IR spectrum from  Cas~A, comprised of the {\it IRAS} (12 -- 100 $\mu$m) and SCUBA observations, from which the synchrotron component has been subtracted, was fitted by Dunne et al. (2003) with a hot (112$^{+11}_{-21}$ ~K) and a cold (18$^{+2.6}_{-4.6}$~K) dust component. The hot dust component detected by the {\it IRAS} has a mass of $\sim 10^{-3}$~M$_{\sun}$, and consists of supernova dust heated by the interaction of the ejecta with the ambient medium (Arendt, Dwek, \& Moseley 1999).  The cold dust has a mass between $\sim$ 2 and 12~M$_{\sun}$, depending on its radiative properties (Dunne et al. 2003 and \S2 below).

The properties and origin of the cold dust are difficult to explain even if one adopts the lower mass estimate.The inferred dust mass is too large to be of either circumstellar or interstellar origin. Such origin will suggest the presence of an associated $\sim$~200~M$_{\sun}$ mass of hydrogen, clearly ruling out the former. An interstellar cloud of that mass interacting with the remnant will give rise to optical and IR emission lines and to dynamical effects that are unobserved. 
Below we argue that the inferred dust mass is also too large to be of supernova origin. Furthermore, the gas associated with the dust, will collisionaly heat the dust to a temperature that is significantly higher than that inferred from the SCUBA observations.  Nevertheless, ruling out an interstellar or circumstellar origin, Dunne et al. (2003) adopted the least implausible explanation, and concluded that the dust must be of SN origin.

This interpretation of the SCUBA observations faces several difficulties:
\begin{enumerate}
\item The required dust condensation efficiency in the SN ejecta is too high even for the amorphous or clumpy dust aggregates. 
Cas~A is believed to be the remnant of the explosion of a 10 to 30~M$_{\sun}$ star (Fesen et al. 2001). The total mass of condensible elements (C, Mg, Si, S, Ca, Fe, Ni, and associated O) ejected in the explosion of 10 and 30~M$_{\sun}$ stars is about 0.3 and  2~M$_{\sun}$, respectively (Woosley \& Weaver 1995). So even the lowest estimate of $\sim$~2~M$_{\sun}$ of cold dust requires a very unlikely condensation efficiency of 100\% in the supernova ejecta of a 30 ~M$_{\sun}$ star. The large dust mass becomes even more embarrassing if, as suggested by the distribution of elemental abundance ratios in the ejecta, the progenitor of Cas~A was only a 12~M$_{\sun}$ star (Willingale et al. 2002). Even allowing for a massive progenitor, the required condensation efficiency is too high considering the fact that important refractory elements (Si, Ca, and Fe) are observed to be in the gas phase of the remnant, and to be more uniformly distributed than required by the Dunne et al. (2003) interpretation of the SCUBA data (Hwang, Holt, \& Petre 2000; Hughes et al. 2000, Willingale et al. 2002). Finally, the total mass of dust that must have formed in the ejecta may actually be greater than implied from the SCUBA observations, since a fraction of it may have been destroyed by thermal sputtering.
\item The density of the gas associated with the cold dust is too high to account for the low dust temperature. Associated with the refractory elements produced in a 30~M$_{\sun}$ star are about 4~M$_{\sun}$ of other abundant metals, primarily oxygen, that are not locked up in dust. 
We can derive a lower limit to the electron density of the emitting region by assuming that it is smoothly distributed between the forward and reverse shocks, and that it covers about 15\% of the sky as seen from the center of the explosion. The volume of the region, for an adopted distance of 3.4~kpc, is then $\sim 2\times10^{56}$ cm$^{-3}$, giving a minimum electron density of $\sim$~3~cm$^{-3}$. Actual electron densities are significantly higher, since oxygen is expected to be multiply ionized in the emitting region. However, even at the low density of 3~cm$^{-3}$,
 typical 0.01 $\mu$m-sized supernova condensates will be collisionally heated to temperatures of about 50~K (Dwek 1987). Put differently, a density of 3~cm$^{-3}$ is about 100 times higher than that required to heat the dust particles to the observed temperature of 18~K.  To achieve this cold dust temperatures at these high electron densities, the radii of the dust particles in the ejecta must be $\gtrsim 1\ \mu$m. However, such radii are much larger than those predicted from nucleation calculations (Kozasa, Hasegawa, \& Nomoto 1991).
 \item Finally, the 170 $\mu$m {\it ISO} observations of Cas~A are actually a detection, implying an even larger amount of colder dust. In their analysis, Dunne et al. (2003) regarded the {\it ISO} 170~$\mu$m observations of Cas~A (Tuffs et al. 1999) as a lower limit, predicting a significantly higher flux from the remnant than observed at that wavelength. However, it was pointed out by the referee that it is difficult to explain how the ISOPHOT instrument on board the {\it ISO} could have missed most of the emission from that region of the remnant. In fact, the observed 170~$\mu$m emission is an actual detection (private communications between the referee and R. J. Tuffs made available to the author), and the 170~$\mu$m flux can be entirely attributed to the hot dust emission. If the 170 $\mu$m flux produced by the cold dust is constrained to be less than 10~Jy, then the SCUBA data can only be fitted 
with $\sim$ 7 to 9~K dust assuming a $\lambda^{-2}$ emissivity law.  The mass of cold dust is then increased to $\sim$ 24 to 16~M$_{\sun}$, respectively, even if it has the high submillimeter radiative efficiency of amorphous or clumpy aggregate material. So regarding the 170 $\mu$m flux as an upper limit on the emission from the cold dust component simply exacerbates the difficulties listed above.
\end{enumerate} 

The problems listed above can be alleviated if the dust responsible for the observed emission consists of conducting needles or whiskers. The high conductivity (low resistivity) of these needles gives rise to a mass absorption efficiency that is several orders of magnitudes greater than that of interstellar graphite or silicates, and even the amorphous clumpy aggregates considered by Dunne et al. (2003). Sufficiently large needles are efficient antenna--like radiators out to millimeter wavelength, depending on their length--to--diameter ratio. Their large radiative efficiency significantly lowers the mass of cold dust required to explain the SCUBA observations. It also increases the density of electrons needed to collisionally heat the dust to the observed temperature. The introduction of such efficient radiators as the source of the observed submillimeter dust emission requires therefore a balance between the need to reduce the mass of the emitting dust without increasing the plasma density beyond the limits that may be imposed from X-ray observations.

Conducting needles have been considered before, in a different context, by Hoyle \& Wickramasinghe (1999) and Wickramasinghe (1992) as a source of starlight opacity that could have produced a non--cosmological cosmic microwave background (CMB) in a steady state cosmology. They were also considered as a way to thermalize starlight and generate the CMB after the big bang in cold big bang models (Wright 1982, Li 2003). It should not be too surprising to find needles in supernova ejecta.
Supernovae condensates may originally be different in morphology or composition from the dust particles commonly found in the diffuse interstellar medium (ISM), and are likely to undergo additional chemical and physical processing in the ISM. There is prior evidence that SN condensates may be different from their ISM counterparts. The hot dust emission from Cas~A detected by the {\it ISO} was interpreted as emission from protosilicate material (Arendt, Dwek, \& Mosleley 1999), and conducting needles may have been detected in SN 1987a (Wickramasinghe \& Wickramasinghe 1993).

In the model proposed here,  the metallic needles were formed in metal-rich ejecta that were expelled at high velocities into the ambient medium. The needles are collisionaly heated by the reverse shock resulting from the interaction of the ejecta with the ambient medium.  In the following section (\S2) we examine the different dust component contributing to the observed IR to submillimeter spectrum of Cas~A, showing that conducting needles are a necessary emission component if one wants to fit the 170~$\mu$m {\it ISO} observations with a reasonable dust mass. In \S3 we calculate the properties of the needles and the electron density required to explain  the SCUBA observations for various type of conducting material. The results of the paper are summarized and discussed in \S4. 

 \section{DUST EMISSION COMPONENTS IN THE SPECTRUM OF CAS A}
 
 Figure 1 depicts the observed infrared (IR) to submillimeter spectrum of Cas~A comprising of the 12, 25, 60, and 100~$\mu$m data obtained by the {\it Infrared Astronomical Satellite} ({\it IRAS}) (Dwek et al. 1987), the 170~$\mu$m datum point obtained by the {\it Infrared Satellite Observatory} ({\it ISO}) satellite (Tuffs et al. 1999), and the 450 and 850~$\mu$m SCUBA data (Dunne et al. 2003). An almost identical 850~$\mu$m flux was reported by Loinard et al. (2003). Also shown in the figure is the synchrotron emission spectrum, extrapolated with a $\nu^{-0.69}$ power law (Mezger et al. 1986). Table 1 summarizes the data used in this paper.

 \begin{figure}
\plotone{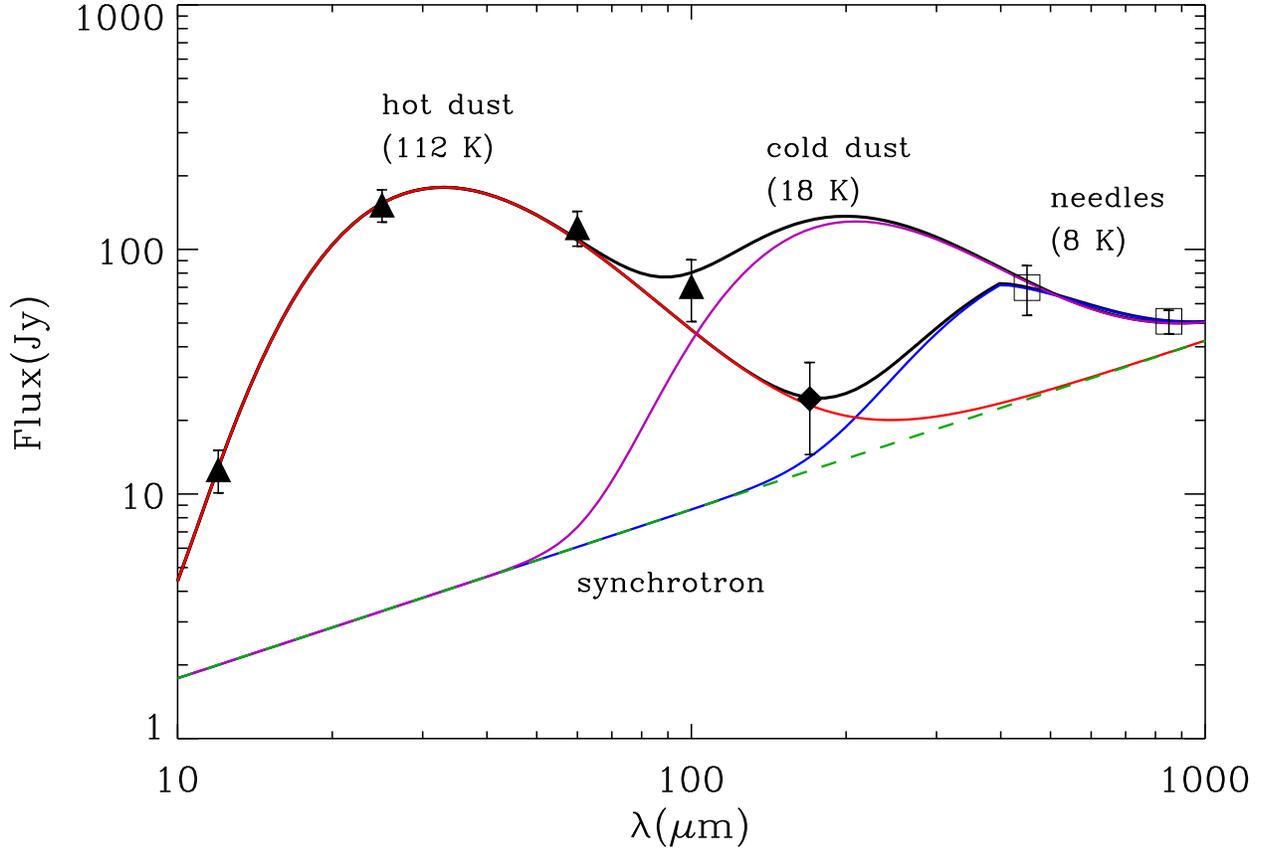}
\caption{The infrared--submillimeter spectrum of Cas~A consisting of the {\it IRAS} 12,  25, 60, and 100~$\mu$m  (filled triangles, Dwek et al. 1987), the {\it ISO} the 170~$\mu$m (filled diamond, Tuffs et al. 1999), and the SCUBA 450 and 850~$\mu$m observations (open squares, Dunne et al. 2003). Regarding the 170~$\mu$m flux as a lower limit, Dunne et al. fitted the data with a hot+cold dust components. Taking the 170~$\mu$m flux as a detection, the data can be fitted with a hot dust+conductive needles components. Needles are required to alleviate the excess mass problem.}
\end{figure}
 
 The specific flux at wavelength $\lambda$ received from an ensemble of dust particles with a total mass $M_d$, located at a distance $D$, and radiating at an equilibrium dust temperature $T_d$, is given by:
 \begin{equation}
F_{\nu}(\lambda) =\left({ M_d\over  D^2}\right)  \kappa(\lambda) B_{\nu}(\lambda,\ T_d) 
\end{equation}
where  $\kappa(\lambda)$ is the mass absorption coefficient of the dust, and $B_{\nu}$ is the Planck function. For a $\kappa(\lambda) \propto \lambda^{-n}$ emissivity law, eq. (1) can be written as:
 \begin{equation}
F_{\nu}(\lambda) =8.30\times 10^9\ { M_d({\rm M}_{\sun}) \over  D_{kpc}^2}  \left[{\kappa(\lambda_0)\over {\rm cm}^2\ {\rm g}^{-1}} \right]\left({\lambda_0\over \lambda}\right)^n \lambda_{\mu{\rm m}}^{-3} \left[ \exp \left({14387.7 \over \lambda_{\mu{\rm m}} T_d}\right)-1\right]^{-1}   \qquad {\rm Jy}
\end{equation}
where $\kappa(\lambda_0)$ is the mass absorption coefficient at some reference wavelength $\lambda_0$

Regarding the {\it ISO} observations as a lower limit, Dunne et al. (2003) fitted the Cas~A spectrum with a hot and cold dust component radiating at an equilibrium temperature of 112~K, and 18~K, respectively. Figure~1 depicts their 2-component fit to the  {\it IRAS} and SCUBA data of Cas~A. 
Interstellar dust particles have a value of $\kappa(450\ \mu{\rm m}) \approx$ 2.5~cm$^2$ g$^{-1}$ (Draine \& Lee 1984). Assuming that the Cas~A dust has similar radiative properties gives a total mass of $\sim 2\times 10^{-3}$~M$_{\sun}$ of hot dust and about 10~M$_{\sun}$ of cold dust. The mass of cold dust may be lower if it has an amorphous or clumpy aggregate structure, for which laboratory measurements and theoretical models give a mass absorption coefficient of $\kappa(450\ \mu{\rm m})$ = 15~cm$^2$ g$^{-1}$ (Ossenkopf \& Henning 1994). The dust mass is then only about 2~M$_{\sun}$, still incompatible with a circumstellar or interstellar origin, but more likely (according to Dunne et al. 2003) to be of supernova origin. 

The model of Dunne et al. faces more difficulties if the {\it ISO} 170~$\mu$m observations are an actual measurement (Tuffs et al. 1999).  The observed flux can then be entirely attributed to the emission from the hot dust component (see Figure 1). A fit to the SCUBA observations would then lead to a colder dust temperature and therefore to a significantly higher dust mass, even if it consists of  amorphous or clumpy aggregates.

The mass of cold dust depends on the mass absorption coefficient, and can be further reduced if the dust is a more efficient emitter at far-IR and submillimeter wavelengths than the amorphous clumpy aggregates considered by Dunne et al. (2003). Conducting needles are just such kind of dust. 

 \section{CONDUCTING NEEDLES}

\subsection{Mass Absorption Coefficient}

Consider a conducting needle--like dust particle represented by a circular cylindrical of radius $a$, length $\ell$ ($a \ll \ell$), mass density $\rho_m$, and mass $m_d = \pi a^2 \ell \rho_m$. Most of the incident radiant energy is absorbed by ohmic dissipation. 
The absorption cross section is given by $\sigma_{abs} = P/S$, where $S = (c/8\pi)E^2$ the time-averaged Poynting vector of the incident radiation, $E$ the amplitude of the incident electric field, and $P$ the power absorbed in the grain, are given by (Wright 1982):
\begin{eqnarray}
P & = & {V^2 \over 2R} = {1\over 3} {(E \ell)^2\over 2R} \\ \nonumber
 & = & {1\over 3} {\pi a^2 (E \ell)^2 \over 2\rho_R \ell}
\end{eqnarray}
In the equation above, $V$ is the voltage on the needle,  $R = \rho_R \ell/\pi a^2$ is its resistance, $\rho_R$ its resistivity, and the factor of $\onethird$ arises from averaging over the angles of incident between the electric field and the needle (Wright 1982). The absorption cross section is $\sigma_{abs} = (4 \pi /3 c) \pi a^2 \ell/\rho_R$, and the mass absorption coefficient is:
\begin{equation}
\kappa_0 \equiv {\sigma_{abs} \over m_d} = {4 \pi \over 3} \left({1\over \rho_m \rho_R c}\right) \qquad.
\end{equation}

The mass absorption coefficient of the needles is nearly constant for wavelengths $\gtrsim 100\ \mu$m (Hoyle \& Wickramasinghe 1999, Wright 1987), with a long--wavelength cutoff at $\lambda_0$ given by (Wright 1982):
\begin{equation}
\lambda_0 = {1\over 2} \rho_R c \ {(\ell/a)^2\over \ln(\ell/a)} \qquad,
\end{equation}
after which $\kappa(\lambda) \propto \lambda^{-2}$ at wavelengths $> \lambda_0$. 

Table 2 lists several physical properties of different conducting and semi-conducting material consisting of elements that are likely to condense out of metal-rich SN ejecta. The electrical resistivity of metals is generally a sum of two independent contributions: (1) the resistivity caused by the collisions of the conduction electrons with lattice phonons, and (2) the resistivity caused by the scattering of the electrons by impurity atoms and lattice defects. The first contribution is strongly temperature dependent, generally dropping by 2 to 3 orders of magnitude from room temperature to 20~K.  For example, the resistivity of pure iron drops from $\sim 10^{-5}\ \Omega$~cm at 293~K to a value of $\sim 3\times 10^{-8}\ \Omega$~cm at 20~K (1 $\Omega$~cm = 1.139$\times 10^{-12}$~sec). The contribution of impurities and lattice defects to the resistivity, also referred to as the residual resistivity of the material, is independent of temperature, and can dominate the resistivity at sufficiently low temperatures. For alloys, such as nichrome (80\% Ni--20\%Cr), the resistivity is virtually constant between temperatures of 300 and 20~K. At very low concentrations, impurities can contribute about  4~$\mu\Omega$~cm per percent of impurity to the resistivity (Kittel 1963).  SN condensates are likely to have embedded impurities, however, their concentration is unknown. Given the uncertainties in the resistivities of some elements at very low temperatures, we will take the value of $\rho_R$ as a variable in our calculation. 

\subsection{Temperature and Long Wavelength Cutoff}

For given material properties, the needle emission spectrum is completely determined by its temperature and long-wavelength cutoff. The observational constraints are the $F_{\nu}(450\ \mu{\rm m})$ to $F_{\nu}(850\ \mu{\rm m})$ flux ratio, which is equal to =3.5$\pm$2, and the rapid drop in the spectrum at 170~$\mu$m. Figure~2 depicts the 450-to-850~$\mu$m (dashed lines) and the 450-to-170~$\mu$m flux ratios (solid lines) as a function of the cutoff wavelength and needle temperature.  
The figure shows that the SCUBA observations are well matched by 8.2~K needles with a long wavelength cutoff of $\lambda_0$ = 400 $\mu$m. Figure 3 depicts the $\ell/a$ ratio for $\lambda_0$ = 400 $\mu$m as a function of needle resistivity. The $\ell/a$ ratio varies from $\sim$ 4000 to about 100 over the range of resistivities depicted in the figure. For nichrome, which has a constant resistivity as a function of temperature, the $\ell/a$ ratio is about 260 giving a needle length of $\sim$ 2.6 $\mu$m for a radius of $a = 0.01\ \mu$m. 
Figure 1 depicts the fit of the needle spectrum to the SCUBA data. The needle model predicts a significantly lower flux at wavelengths below $\sim$ 300~$\mu$m compared to the cold dust model of Dunne et al. (2003). The 100 to 400 $\mu$m flux from Cas~A is therefore an important discriminator between the cold dust and the needle model. 

 \begin{figure}
\plotone{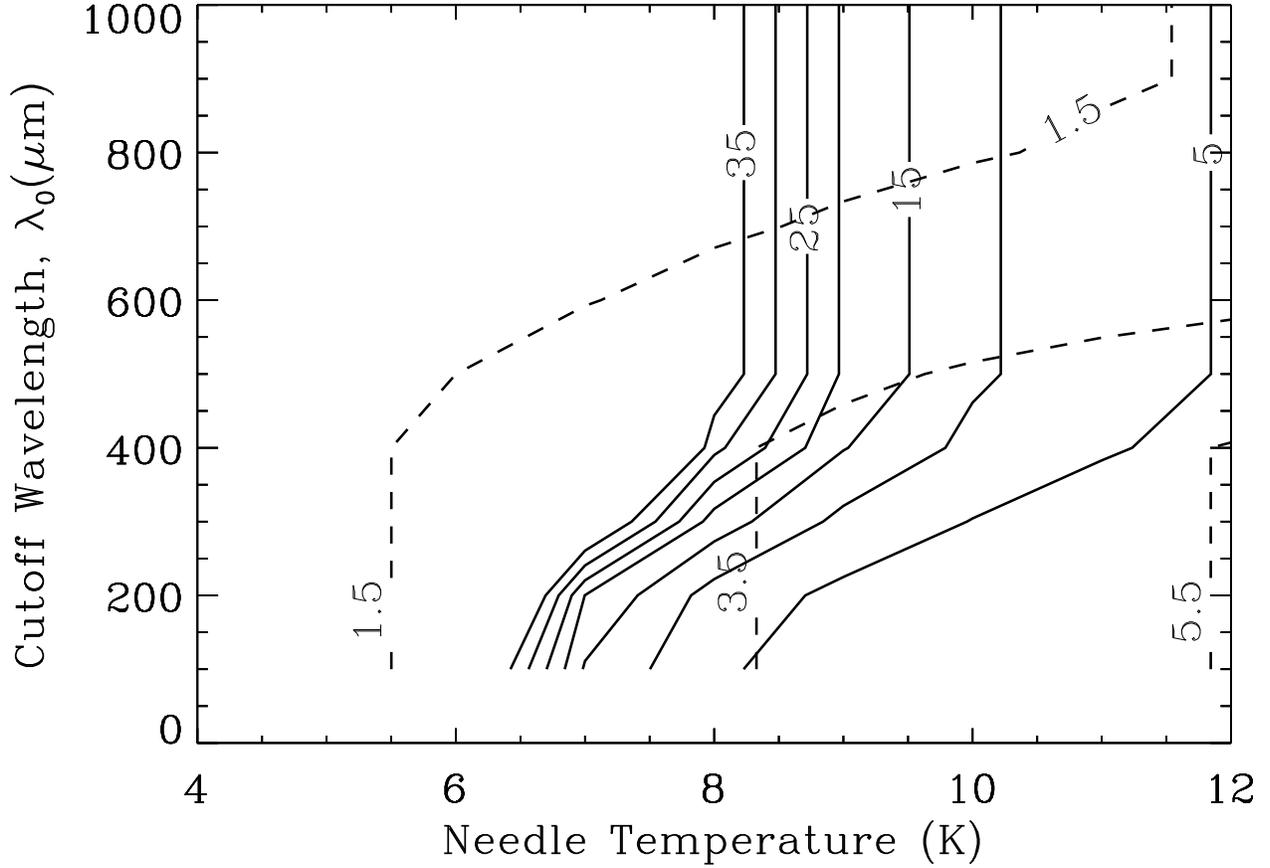}
\caption{Contour levels of the $R_1\equiv F_{\nu}(450\ \mu{\rm m})/F_{\nu}(850\ \mu{\rm m})$ (dashed lines) and the $R_2 \equiv F_{\nu}(450\ \mu{\rm m})/F_{\nu}(170\ \mu{\rm m})$ (solid lines) flux ratios are plotted versus the cutoff wavelength, $\lambda_0$, and needle temperature. A good fit to the observational constraints $R_1=3.5\pm2$ and $R_2 >> 1$ require $\lambda_0$ = 400 $\mu$m, and a needle temperature of 8.2~K. The needle spectrum is depicted in Figure 1.} 
\end{figure}

\subsection{Total Mass}

The total mass ${\cal M}_d$, of needles radiating at temperature $T_d$ is given by: 
\begin{equation}
{\cal M}_d = { D^2 S_{\nu}(\lambda) \over \kappa(\lambda) B_{\nu}(\lambda,\ T_d)} \qquad ,
\end{equation}
where $S_{\nu}(\lambda)$ is the observed flux density, and $D$=3.4~kpc is the distance to the remnant. Figure 3 shows the mass of dust required to produce a 450~$\mu$m flux of 45.4~Jy for a needle temperature of 8.2~K. An average material density of 8.15~g~cm$^{-3}$ was adopted in these calculations. Unlike the cold dust model, the mass of elements locked up in needles is only a small fraction of the total mass of metals produced in the explosion. However, the masses shown in the figure are only lower limits on the mass of dust that condensed out of the ejecta, since the needles are destroyed by thermal sputtering in the shock-heated gas. In the following we calculate the density of the gas required to heat the needles to a temperature of $\sim$~8~K, and the corresponding rate of grain destruction.  

 \subsection{The Heating and Cooling of Needles}

The heating rate of a needle due to electronic collisions is given by:
\begin{eqnarray}
H_d(T) & = & 2 \pi a \ell n_e (1-f_r) \int g(E) v(E) E_{dep}(E) {\rm d}E \nl \nonumber
& = & 2 m_d n_e (1-f_r) \int g(E) v(E) \left[{E_{dep}(E)\over \rho_m a}\right] {\rm d}E
\end{eqnarray}
where $g(E) = 2 \pi^{-1/2} (kT)^{-3/2} E^{1/2} \exp(-E/kT)$ is the Maxwell distribution function of electron energies $E$ at temperature $T$, $v(E)$ and $n_e$ are, respectively, the electron velocity and number density, $f_r$ is the fraction of incident electrons that are reflected from the grain surface, and $E_{dep}(E)$ is the energy deposited by incident electrons with energy $E$ in the needle. If the electrons are stopped in the grain, $E_{dep} = E$, otherwise $E_{dep}=\bar x  ({\rm d}E/{\rm d}x)$, where $\bar x \approx 2a$ is an average distance traveled by an electron through the needle, and (d$E$/d$x$) is the electronic stopping power (the energy lost in the solid per unit length traversed by the electrons). 

The range of electrons with energies between 370 and 10$^4$ eV is given by (Iskef, Cunningham, \& Watt 1983):
\begin{equation}
R_e(\mu{\rm m}) = 2.59\times10^{-6}\ \left[{\rho_m \over {\rm g\ cm}^{-3}}\right]^{-1}\ E(eV)^{1.492}
\end{equation}
%
Taking $R_e = 2 a$ = 0.02 $\mu$m as the effective pathlength through the cylinder, we get that the maximum energy $E_*$ at which the electrons are stopped in the needle is given by:
\begin{eqnarray}
E_*(eV) & = & 403 \left[{\rho_m \over {\rm g\ cm}^{-3}}\right]^{0.67} \nl \nonumber
 & \approx & 1.64\times 10^3 \qquad {\rm for}\ \rho_m = 8.15\ {\rm gr\ cm}^{-3}
\end{eqnarray}

\noindent
Differentiating eq. (8) w.r.t. $E$ gives the electronic stopping power in the 370 to 10$^4$ eV energy range:
\begin{equation}
{{\rm d}E\over {\rm d}x }  =  2.59\times10^9 \left({\rho_m\over  {\rm g\ cm}^{-3}}\right) \left({E\over {\rm eV}}\right)^{-0.492}\qquad {\rm eV\ cm}^{-1}  \qquad,
\end{equation}
The dust heating rate can be now written as the sum:
\begin{eqnarray}
H_d(T) & = & 2 m_d n_e (1-f_r) \{\ \int_0^{E_*} g(E) v(E) \left({E\over \rho_m a}\right) {\rm d}E \nl \nonumber
    &  & + \int_{E_*}^{\infty} g(E) v(E)\ {1\over \rho_m} \left({ {\rm d}E\over {\rm d}x}\right) {\rm d}E\  \}
\end{eqnarray}

Radiating at temperature $T_d$, the needle cools at a rate given by:
\begin{equation}
 L_d =  4 m_d\ \int \kappa(\nu_0) \pi B_{\nu}(\nu,\ T_d) \ {\rm d}\nu 
 \end{equation}
 where 
 \begin{eqnarray}
 \kappa({\lambda}) & = & \kappa_0 \qquad \qquad  \qquad {\rm for}\ \lambda < \lambda_0 \nl
  & = & \kappa_0\ \left({\lambda_0\over \lambda}\right)^{-2} \qquad {\rm for}\ \lambda \geq \lambda_0
\end{eqnarray} 
where $\kappa_0$ and $\lambda_0$ are given by eqs. (4) and (5), respectively. 

In equilibrium, the dust cooling rate equals its heating rate by electronic collisions. Figure 4 depicts the electron density $n_e$ as a function of needle resistivity at various fixed plasma temperatures, represented by the diagonal lines in the figure. The figure depicts two plasma components identified in the X-ray analysis of Cas~A (Willingale et al. 2003): a hot component characterized by  \{$n_e,\ kT_e$\} =  \{16$\pm$3 cm$^{-3}$, 3.27$\pm$0.86 keV\}, and a cool component with \{$n_e,\ kT_e$\} = \{61$\pm$15 cm$^{-3}$, 0.45$\pm$0.10 keV\}. The shaded areas in the figure represent the allowable density and temperature combinations for the two plasma components, defined by the 1$\sigma$ uncertainties around their central values. In the hot plasma, the needles are mostly transparent to the incident electrons. The dust heating rate is then independent of electron temperature (Dwek 1987), accounting for the narrowness of the region between the two temperatures that bound the hot plasma. The figure shows that the needle resistivity needs to be between $\sim~(2-5)\times10^{-6}\ \Omega$ cm, and could reside in either the hot or cool  (or both) plasma components of the remnant.
 
\begin{figure}
\plotone{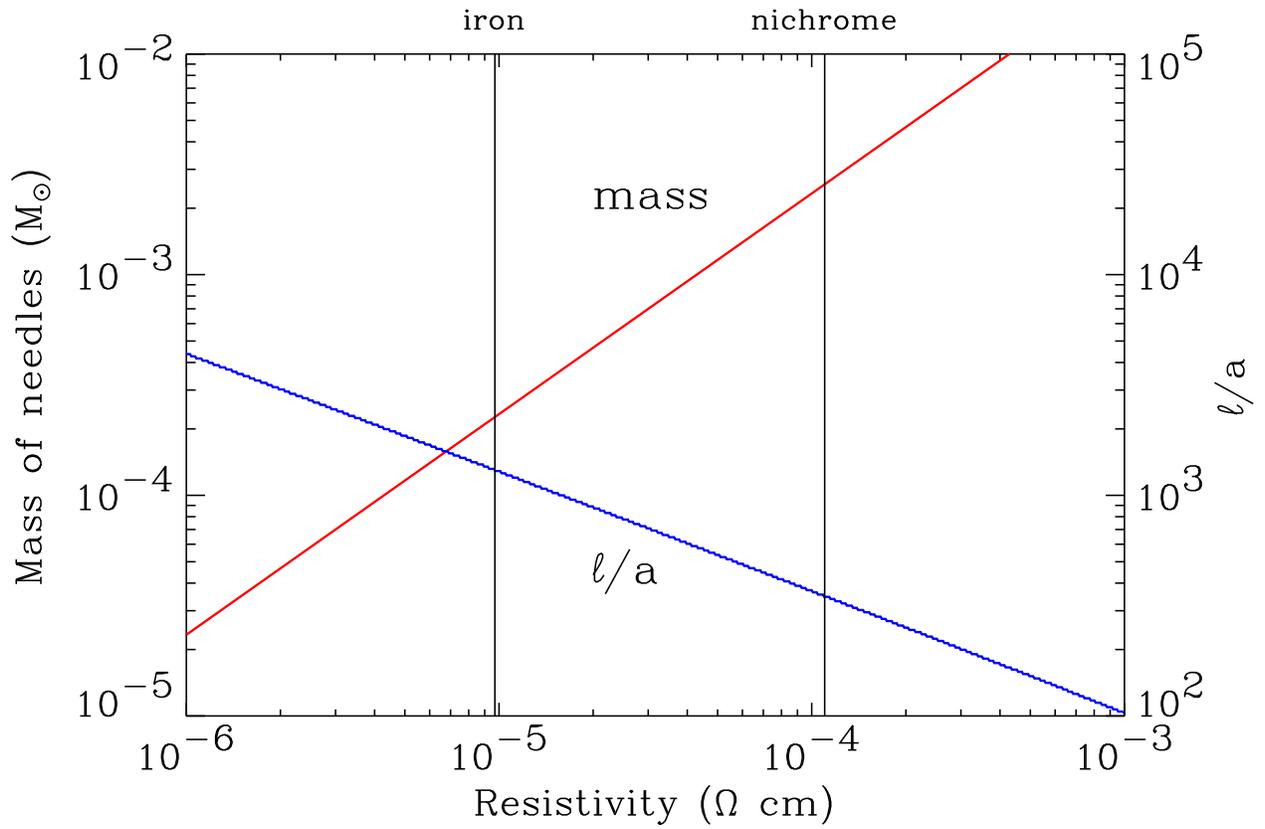}
\caption{Plot of the mass of 8.2~K needles required to fit the SCUBA data (see Figure 1), and the needle length-to-radius, $\ell/a$, ratio as a function of needle resistivity. Also shown in the figure are the room temperature resistivities of iron (Fe), and nichrome (Ni-Cr).}
\end{figure}

\begin{figure}
\plotone{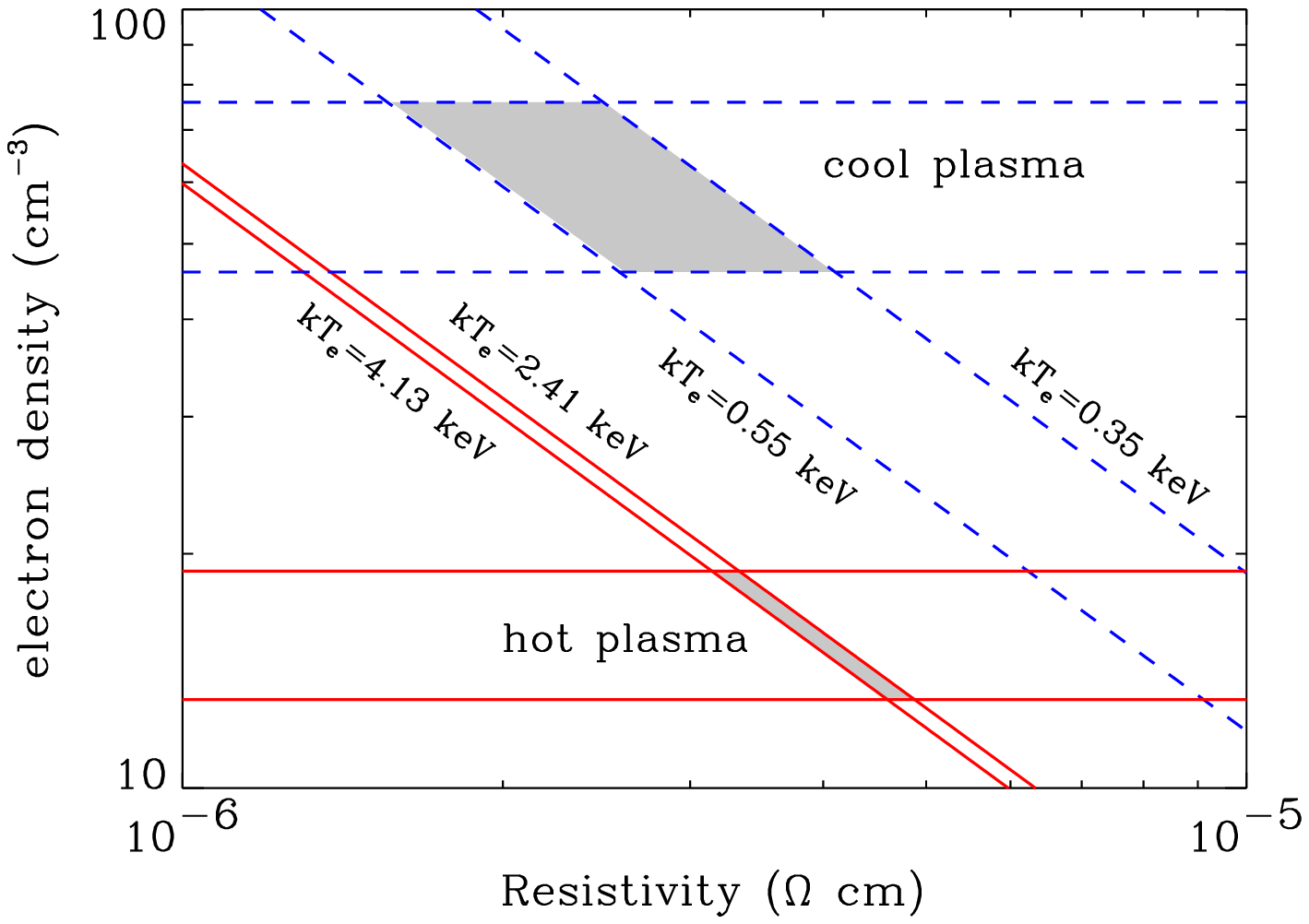}
\caption{The electron density required to heat a needle to 8.2~K is plotted versus needle resistivity for different plasma temperatures. The shaded areas define the parameters of the cool and hot plasma components in Cas~A, defined by \{$n_e,\ T_e$\} = \{61$\pm$15 cm$^{-3}$, 0.45$\pm$0.10 keV\}, and \{16$\pm$3 cm$^{-3}$, 3.27$\pm$0.86 keV\}, respectively.}
\end{figure}

\subsection{Needle Destruction by Sputtering}
 Dust particles in a shocked plasma are eroded by thermal sputtering. For plasma temperatures of $\sim 10^7$~K ($\sim$ 0.86 keV), the lifetime of silicate dust against sputtering by heavy nuclei, primarily oxygen, in the SN ejecta is given by (Dwek, Foster, \& Vancura 1996):
 \begin{equation}
\tau({\rm yr}) \approx 5\times 10^5\ {a(\mu{\rm m})\over n({\rm O})}
\end{equation}
where $n$(O) is the number density (in cm$^{-3}$) of the oxygen nuclei in the hot gas. Iron dust has a lower surface binding energy of about 4.3~eV, compared to the surface binding energy of silicates, which is about 6~eV (Dwek \& Scalo 1980). Iron dust will therefore have a shorter lifetime than silicate particles of identical size. To be definite, we adopt a value of $10^5$ for the numerical coefficient in the expression for the dust lifetime. 
Assuming that the density of heavy elements is approximately given by the electron density in the two different plasma components, and taking $a = 0.01\ \mu$m, we get that the lifetime of iron needles is about 60~yr in the cool, and about 16~yr in the hot plasma component. Both times are shorter than the range of ages of these two plasma components which are, respectively, about 80 (20--273) and 131 (101-182) yr (Willingale et al. 2003). The radiating dust mass represents therefore the fraction of the total mass of shocked dust which dust has not yet been destroyed. A rough estimate for the actual mass of needles that must have been condensed in the ejecta in order to produce the observed emission can be obtained by multiplying the masses given in Figure 3 by the ratio of the age of the plasma to the sputtering lifetime. Typical needle masses need therefore to be 2 to 10 times larger than indicated in the figure if they reside, respectively, in the cool or hot plasma component of the remnant. Adopting average resistivities for the needles in the two plasma components (Figure 4), we get that the total mass of needles required to produce the SCUBA observations is $\sim \times10^{-4}$~M$_{\sun}$ if they reside in the cool plasma, and  $\sim \times10^{-3}$~M$_{\sun}$ if they reside in the hot plasma. There may yet be an additional reservoir of colder dust. So even after the effects of grain destruction have been taken into account, the total mass of needles required to have formed in the ejecta still represents a very small fraction of the mass of available metals in the ejecta.

\section{SUMMARY AND DISCUSSION}
Recent SCUBA observations of Cas~A (Dunne et al. 2003, Loinard et al. 2003) have revealed an excess of submillimeter emission above the extrapolated synchrotron emission from the remnant. Dunne et al. (2003) interpreted the excess as emission from about 2~M$_{\sun}$ of cold, 18~K, dust that has formed in the SN ejecta. This dust mass requires the implausible scenario that essentially {\it all} the condensible elements that formed in ejecta be locked up in the dust. X-ray observations rule out such scenario. Their interpretation faces more difficulties if grain destruction by sputtering is taken into account, and if the 170~$\mu$m observations of the remnant are also included as a detection in the analysis. The cold dust temperature then becomes $\sim$ 8~K, increasing the inferred cold dust mass to about 20~M$_{\sun}$.

A conductive needle origin for the cold dust emission from Cas~A provides a simpler alternative explanation than the one put forward by Dunne et al. (2003), even when the 170~$\mu$m observations are included in the analysis. The results of the paper can be briefly summarized as follows:
\begin{enumerate}
\item the submillimeter spectrum can be fit with 8.2~K needles having a long wavelength cutoff of 400~$\mu$m (Figures 1 and 2),
\item adopting a range of $10^{-6}$ to $10^{-3}\ \Omega$~cm of material resistivities, the mass of needles required to produce the SCUBA observations falls between 2$\times10^{-5}$ and 0.02~M$_{\sun}$, respectively, and that the needle length-to-radius ($\ell/a$) ratio is between 4000 and 100, respectively (Figure 3),
\item balancing the radiative cooling rate of the needles at 8.2~K to the collisional heating rate for a range of plasma densities and temperatures, the range of needle resistivities that are compatible with observational X-ray constraints can be narrowed to (2--5)$\times10^{-6}\ \Omega$~cm (Figure 4). These values suggest that the needles are made of pure conducting material (iron?) that includes about 0.5 to 1\% of embedded impurities
\item the needles are destroyed by sputtering in the hot gas. The needle lifetime is shorter than the ionization age of the various X-ray emitting components, suggesting a continuous supply of newly swept-up dust. The mass of needles required to produce the SCUBA observations is therefore increased to $\sim10^{-4}$~M$_{\sun}$ if they reside in the cool plasma, and  $\sim10^{-3}$~M$_{\sun}$ if they reside in the hot plasma. There may be an additional reservoir of cold dust that has not been shocked yet.    
\end{enumerate}
 
The needle model proposed here is consistent with a scenario in which the metallic needles are formed in bullets of metal--rich ejecta that were expelled at high velocities into the ambient medium. Such scenario is similar to the one observed in SN~1987A, in which the increased ionization state of its envelope, the early appearance of X- and $\gamma$-rays from the radioactive decay of $^{56}$Co, and the width of the spectral lines of heavy elements, suggested that metal--rich material from the inner region of the ejecta was macroscopically mixed and expelled at velocities of $\sim$ 3000 km s$^{-1}$ (Graham 1988, Arnett et al. 1989).  
The same picture emerges from optical and X--ray studies of the ejecta of Cas~A. The spatial distribution of the nucleosynthesis products in the remnant shows that the most Fe--rich material is very clumpy and located at the outer edge of the ejecta (Hughes et al. 2000, Willingale et al. 2002), suggesting that the stellar layers were macroscopically mixed, spatially inverted, and ejected as bullets into the ambient medium in an asymmetric explosion. 
The interaction of the ejecta with the ambient medium will drive a reverse shock wave through the ejecta. The shocked plasma will then collisionally heat the dust to the observed temperature. 
The submillimeter emission detected by the SCUBA from Kepler (Morgan et al. 2003) may have a similar origin. 

Conductive needles may be efficiently aligned in a magnetic field, so that the cold dust emission may be partially polarized, depending on the geometry of the magnetic field. The submillimeter polarization from the dust can dominate that from the synchrotron emission. The detection of polarization at wavelengths $\lesssim 450 \mu$m, and a drop in the cold dust spectrum from 400 to 170~$\mu$m will therefore be definitive confirmations of the proposed model.
  
The results of our analysis show that the SCUBA observations need not imply the detection of massive amounts of dust in the remnant. Conductive needles offer a better explanation with only a fraction of the mass inferred by Dunne et al. (2003) and Morgan et al. (2003). Nevertheless, supernovae may yet be important sources of interstellar dust. Here we argue that the evidence put forward sofar is still inconclusive. 

Acknowledgements: In the course of this work I have benefitted from useful discussions with Kazik Borkowski, Eric Gotthelf, Una Hwang,  Harvey Moseley, and Ned Wright.  Dale Fixsen and David Cottingham steered me to useful sources of resistivity data. I thank Haley Morgan for useful communications, and Rick Arendt for many insightful  discussions and comments on the manuscript. I am especially grateful to the referee, Jean-Philippe Bernard, for suggesting the inclusion of a spectral fit to the data in the manuscript, for pointing out the importance of the 170~$\mu$m datum point, and for his thorough review and detailed comments that resulted in a significantly improved presentation of the model. ED acknowledges NASA's Astrophysics Theory Program for support of this work.

\clearpage

\begin{deluxetable}{lccccccc}
\tablecaption{Observed Infrared and Submillimeter Fluxes From Cas~A}
\tablehead{
\colhead{Emission component} &
 \colhead{ 12 $\mu$m} &
 \colhead{ 25 $\mu$m} &
 \colhead{ 60 $\mu$m} &
 \colhead{ 100 $\mu$m} &
 \colhead{ 170 $\mu$m} &
 \colhead{ 450 $\mu$m} &
  \colhead{ 850 $\mu$m}
  }
 \startdata 
Total$^1$ & 16.9 $\pm$2.5 & 152$\pm$23 & 123$\pm$20 & 71$\pm$20 & 35$\pm$10 & 69.8$\pm$16.1 & 50.8$\pm$5.6 \nl
Synchrotron$^2$ & 2.0 & 3.3 & 6.1 &8.7 & 12.5 & 24.4 & 37.8  \nl
Ionic lines & 4.3 & $<$ 2 & $<$ 15 & $<$ 1 & 11.5 & \nodata & \nodata \nl
 Thermal dust & 10.6$\pm$2.5 & 149$\pm$23 & 117$\pm$20 & 62$\pm$20 & 11$\pm$10 & 45.4$\pm$16.1 & 13.0$\pm$5.6 \nl
 \enddata
\tablenotetext{1}{Fluxes in units of Jy. {\it IRAS} 12, 25, 60, and 100 $\mu$m data are from Dwek et al. (1987), 170~$\mu$m flux from Tuffs et al. (1999), and SCUBA 450 and 850 $\mu$m fluxes are from Dunne et al. (2003).}
 \tablenotetext{2}{Synchrotron fluxes calculated for a $\nu^{-0.69}$ power law (Mezger et al. 1986), normalized to 2.0~Jy at 12~$\mu$m.}
\end{deluxetable}


\begin{deluxetable}{lccccc}
\tabletypesize{\footnotesize}
\tablecaption{Summary of Physical Properties of Different Materials$^1$}
\tablehead{
\colhead { } &
\colhead { } &
\colhead{ Pure iron} &
\colhead{Nichrome} &
\colhead{Silicon carbide} &
\colhead{ Graphite} \nl
\colhead { } &
\colhead { } &
\colhead{(Fe)} &
\colhead{(Ni-Cr)} &
\colhead{(SiC)} &
\colhead{(C)} 
}
\startdata
Mass density, $\rho_m$(g cm$^{-3}$) &  & 7.8 & 8.5 & 3.2 &2.2 \nl
Resistivity, $\rho_R (\Omega$ cm)  &  & 9.7$\times10^{-6}$ & 1.1$\times10^{-4}$ &  2.0$\times10^{-4}$ & 9.1$\times10^{-4}$ \nl 
Mass absorption coefficient, $\kappa_0$(cm$^2$ g$^{-1}$) & & 1.3$\times10^6$ & 1.0$\times10^5$ & 1.5$\times10^5$ & 4.8$\times10^4$ \nl
\enddata
\tablenotetext{1}{Resistivities measured at 293~K.  Data taken from the CRC handbook of Chemistry and Physics, 84th Edition, 2003--2004. For unit conversion: 1 $\Omega$ cm = 1.139$\times10^{-12}$ s. The value of $\kappa_0$ was calculated using eq. (4).} 
\end{deluxetable}


\begin{references}
\reference{ } Arendt, R. G. 1989, \apjs, 70, 181
\reference{ } Arendt, R. G., Dwek, E., \& Moseley, S. H. 1999, \apj, 521, 234
\reference{ } Arnett, D. W., Bahcall, J. N., Kirshner, R. P., \& Woosley, S. E. 1989, \araa, 27, 629
\reference{ } Braun, R. 1987, \aap, 171, 233
\reference{ } Dinerstein, H. L., Capps, R. W., Dwek, E., \& Werner, M. W. 1982, \apj, 255, 552
\reference{ } Dunne, L., Eales, S., Ivison, R., Morgan, H., \& Edmunds, M. 2003, Nature, 424, 285
\reference{ } Draine, B. T., \& Lee, H. M. 1984, \apj, 285, 89
\reference{ } Dwek, E, \& Scalo, J. M. 1980, \apj, 239, 193
\reference{ } Dwek, E., \& Werner, M. W. 1981, \apj, 248, 138
\reference{ } Dwek, E. 1987, \apj, 322, 812
\reference{ } Dwek, E., Dinerstein, H. L., Gillett, F. C., Hauser, M. G., \& Rice, W. L. 1987, \apj, 315, 571
\reference{ } Dwek, E., Foster, S. M., \& Vancura, O. 1996, \apj, 457, 244
\reference{ } Dwek, E. 1998, \apj, 501, 643
\reference{ } Fesen, R. A., Morse, J. A., Chevalier, R. A., Borkowski, K. J., Gerardy, C. L., Lawrence, S. S., \& van den Bergh, S. 2001, \aj, 122, 2644
\reference{ } Graham, J. R. 1988, \apj, 335, L53
\reference{ } Gotthelf, E. V., Koralesky, B., Rudnick, L., Jones, T. W., Hwang, U., \& Petre, R. 2001, \apj, 552, L39
\reference{ } Hoyle, F., \& Wickramasinghe, N. C. 1999, \apss, 268, 77
\reference{ } Hughes, J. P., Rakowski, C. E., Burrows, D. N., \& Slane, P. O. 2000, \apj, 528, L109
\reference{ } Hwang, U., Holt, S. S., \& Petre, R. 2000, \apj, 537, L119
\reference{ } Iskef, H., Cunningham, J. W., \& Watt, D. E. 1983, Phys. Med. Biol., 28, 535
\reference{ } Jones, A. P. 2000, JGR, 105, 10257
\reference{ } Kittel, C. 1963, "Quantum Theory of Solids", (Wiley: New York)
\reference{ } Kozasa, T, Hasegawa, H., \& Nomoto, K. 1991, \aap, 249, 474
\reference{ } Lagage, P. O., Claret, A., Ballet, J., Boulanger, F., C\'esarsky, C. J., C\'esarsky, D., Fransson, C., \& Pollock, A. 1996, \aap, 315, L273 
\reference{ } Li, A. 2003, \apj, 584, 593
\reference{ } Loinard, L., Lequeux, J., Tilanus, R. P. T., \& Lagage, P. O. 2003, RevMexAA, 15, 267
\reference{ } Mezger, P. G., Tuffs, R. J., Chini, R., Kreysa, E., \& Gem\"und, H.-P. 1986, \aap, 167, 145
\reference{ } Morgan, H. L., Dunne, L., Eales, S. A., Ivison, R. J., \& Edmunds, M. G. 2003, \apj, 597, L33
\reference{ } Ossenkopf, V., \& Henning, Th. 1994, \aap, 291, 943 
\reference{ } Saken, J. M., Fesen, R. A., \& Shull, J. M. 1992, \apjs, 81, 715
\reference{ } Todini, P., \& Ferrara, A. 2001, \mnras, 325, 164 
\reference{ } Tuffs, R. J., Fischera, J., Drury, L. O'C., Gabriel, C., Heinrichsen, I., Rasmussen, I. \& V\"olk, H. J., 1999,
in  The Universe as seen by {\it ISO}, (ESA SP-427), eds P. Cox \& M. F. Kessler (ESA: Nordwijk), p. 241
\reference{ } Wickramasinghe, N. C. 1992, \apss, 198, 161
\reference{ } Wickramasinghe, N. C., \& Wickramasinghe, A. N., 1993, \apss, 200, 145
\reference{ } Willingale, R., Bleeker, J. A. M., van der Heyden, K. J., Kaastra, J. S., \& Vink, J.  2002, \aap, 381, 1039
\reference{ } Willingale, R., Bleeker, J. A. M., van der Heyden, K. J., \& Kaastra, J. S. 2003, \aap, 398, 1021
\reference{ } Woosley, S. E., \& Weaver, T. A. 1995, \apjs, 101, 181
\reference{ } Wright, E. L., Harper, D. A., Lowenstein, R. F., Keene, J., \& Whitcomb, S. E. 1980, \apj, 240, L157
\reference{ } Wright, E. L. 1982, \apj, 255, 401
\reference{ } Wright, E. L. 1987, \apj, 320, 818
\reference{ } 
\reference{ } 
\reference{ } 
\reference{ } 
\reference{ } 

\end{references}
\end{document}